\def\aa{{A\&A}}

\def\aj{{AJ}}
\def\annrev{{ARA\&A}}
\def\apj{{ApJ}}

\def\mnras{{MNRAS}}
\def\nat{{Nature}}

\def\prl{{Phys. Rev. Lett.}}

\def\amin{$^\prime$}
\def\farcm{\hbox{$.\mkern-4mu^\prime$}}

\documentclass[cup5b]{caps}
\usepackage{graphicx}
\usepackage{amssymb}
\usepackage{ociwsymp3}
\HeadText{M. Birkinshaw}

\begin{document}

\pagenumbering{arabic}

\author[]{MARK BIRKINSHAW\\
          Department of Physics, University of Bristol}

\chapter{Using the Sunyaev-Zel'dovich Effect\\
         to Probe the Gas in Clusters}

\begin{abstract}
The thermal Sunyaev-Zel'dovich effect is an important probe of 
clusters of galaxies, and has the attractive property of being
proportional to the thermal energy content of the intracluster 
medium. With the assistance of X-ray data, the effect can be used to
measure the number of hot electrons in clusters, and thus
measure cluster baryon contents. Cluster absolute distances and
other structural parameters can also be measured by combining thermal
Sunyaev-Zel'dovich, X-ray, and other data. This review presents an
introduction to the effect, shows some representative results, and
sketches imminent developments.
\end{abstract}

\section{Introduction}

Ever since the cosmic microwave background radiation was discovered
and interpreted as a thermal signal coming from the epoch of
decoupling (Dicke et al. 1965; Penzias \& Wilson 1965), it has been
seen as a major cosmological tool. High-quality information about the
background radiation is now available. The {\it COBE} mission demonstrated
that the spectrum of the microwave background radiation is precisely
thermal, with a temperature $T_{\rm rad} = 2.728 \pm 0.002 \ \rm K$
(where necessary, all errors have been converted to $\pm 1\sigma$),
and a maximum distortion characterized by a Comptonization parameter
(\S~\ref{sec:thermal}) $\overline{y} < 1.5 \times 10^{-5}$ or a 
chemical potential $|\mu| < 9 \times 10^{-5}$ (Fixsen et al.~1996).
{\it COBE} also demonstrated that the background contains small brightness
fluctuations induced by density perturbations at decoupling with the
character expected from simple inflation models (Gorski et al.~1996;
Hinshaw et al.~1996; Wright et al.~1996).

More recently, {\it WMAP} has measured the power spectrum of these
fluctuations over a wide range of multipole orders, $l$, and confirmed
the existence of the peak in the power spectrum at $l \approx 200$, which
the BOOMERanG (de~Bernardis et al.~2000) and MAXIMA (Hanany et
al.~2000) experiments used to show that the Universe is
close to flat (Jaffe et al.~2001). Fits to the {\it WMAP} power spectrum
have determined several cosmological parameters with good accuracy,
notably the density parameter in baryons, $\Omega_{\rm b} = 0.047 \pm
0.006$, the density parameter in matter, $\Omega_{\rm m} = 0.29  \pm
0.06 $, and the total density parameter $\Omega_{\rm total} = 1.02 \pm
0.02$ (Bennett et al.~2003; Spergel et al.~2003). This review will
adopt a value  $72 \pm 2 \ \rm km \, s^{-1}$ for the Hubble constant,
derived from the {\it WMAP} results and the Hubble Key Project (Freedman et 
al.  2001), and will
ignore the contribution of neutrinos to the Universe's dynamics since
the {\it WMAP} results suggest that $\Omega_\nu$ is negligible.

Additional brightness structures are induced in the microwave
background radiation by massive objects between the epoch of
decoupling and the present. The most important of these brightness
structures is the thermal Sunyaev-Zel'dovich effect (Sunyaev \&
Zel'dovich 1972), which arises from the inverse-Compton scattering of
the microwave background radiation by hot electrons in cluster
atmospheres. Several other, but lower-amplitude, structures are also
induced by clusters. Recent attention has focussed on those generated
by the kinematic Sunyaev-Zel'dovich effect and the Rees-Sciama effect
(Rees \& Sciama 1968), particularly in the version that arises from
the motions of clusters of galaxies across the line of sight (Pyne \&
Birkinshaw 1993; Molnar \& Birkinshaw 2000).

The thermal Sunyaev-Zel'dovich effect was first detected at a high
level of significance in 1978 (Birkinshaw, Gull, \& Northover
1978). As instrumentation developed through the 1980's and 1990's
it became a routine matter to measure the Sunyaev-Zel'dovich effects
of the richest clusters of galaxies (e.g., Myers et al.~1997).
Nevertheless, the use of the effect to explore the physics of the
intracluster medium is still in its infancy because measurements 
remain slow. This has often led to reports about the effect, from
individual clusters, rather than papers giving measurements for
substantial cluster samples. With the development of dedicated
telescopes the situation is now changing, and the effect is becoming
an important feature of cluster studies.

In this review I describe the physics underlying the
Sunyaev-Zel'dovich effect (\S~\ref{sec:physics}), and then show
how measurements of the effect can be used to extract quantities of
physical interest about the clusters and the gas that they contain in
a relatively model-independent way (\S~\ref{sec:uses}) before
describing the methods used to measure the effect and their
limitations and future prospects
(\S~\ref{sec:instruments}). There have been several 
reviews of the Sunyaev-Zel'dovich effect in recent years (Rephaeli
1995a; Birkinshaw 1999; Carlstrom, Holder, \& Reese 2002), and these
should be read to gain a more complete picture of Sunyaev-Zel'dovich
effect research.

\begin{figure*}[t]
\includegraphics[width=1.00\columnwidth,angle=0,clip]{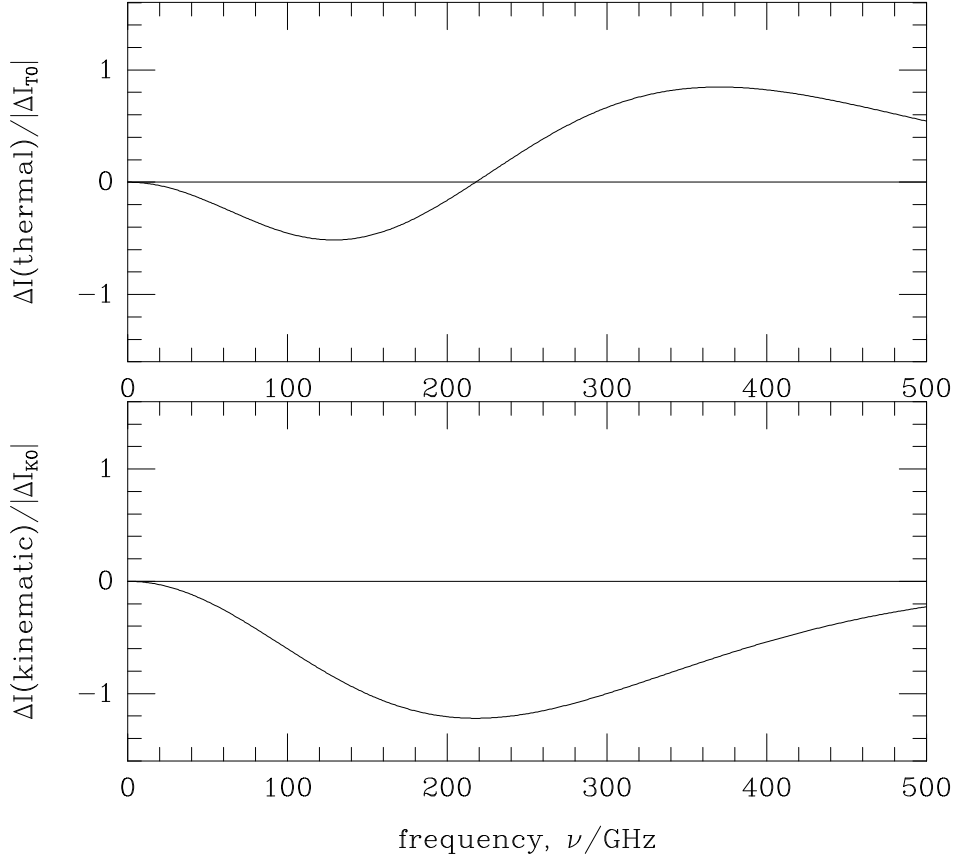}
\vskip 0pt \caption{
The thermal (upper plot) and kinematic (lower plot)
Sunyaev-Zel'dovich effects as functions of frequency, expressed in
terms of the specific intensity change, $\Delta I_\nu$, that they
produce. The largest negative and positive intensity changes in the
thermal effect occur at $129$ and $370 \ \rm GHz$, while the zero of
the thermal effect and the largest negative kinematic effect occur
at $218 \ \rm GHz$.
\label{fig:spectra}}
\end{figure*}

\section{The Physics of the Sunyaev-Zel'dovich Effect}\label{sec:physics}

\subsection{The Thermal Sunyaev-Zel'dovich Effect}\label{sec:thermal}

The Sunyaev-Zel'dovich effects arise because hot electrons in the
atmospheres of clusters of galaxies provide a significant optical
depth to microwave background photons. A cluster of galaxies
containing an X-ray emitting atmosphere with a electron temperature
$T_{\rm e} = 7 \times 10^7 \ \rm K$ ($k_{\rm B} T_{\rm e} = 6 \ \rm
keV$) and average electron density $\overline{n_{\rm e}} = 10^3 \ \rm
m^{-3}$ will have an optical depth for inverse-Compton scattering of
\begin{equation}
 \tau_{\rm e} = \overline{n_{\rm e}} \, \sigma_{\rm T} \, L 
              \approx 10^{-2} \quad ,
\end{equation}
where $\sigma_{\rm T} = 6.65 \times 10^{-29} \ \rm m^2$ is the Thomson
scattering cross-section, for a path length $L \approx 1 \ \rm Mpc$
through the cluster. Since electrons in the intracluster medium
have higher mean energies than the photons, the 1\% of the photons
that are scattered gain energy on average. The mean fractional
frequency change in a scattering is
\begin{equation}
  {\Delta \nu \over \nu} = {k_{\rm B} T_{\rm e} \over m_{\rm e} c^2}
  \approx 10^{-2} \quad ,
\end{equation}
so that the fractional change in the specific intensity of the
microwave background radiation, as seen through the cluster, relative
to directions far from the cluster, is 
\begin{equation}
  {\Delta I_\nu \over I_\nu} \propto \tau_{\rm e} \, {\Delta \nu \over
  \nu} \sim 10^{-4} \quad .
\end{equation}
In the Rayleigh-Jeans part of the microwave background spectrum, where 
$I_\nu \propto \nu^2$, the constant of proportionality is 
$-2$. The full spectrum of the thermal effect is shown in the upper
panel of Figure~\ref{fig:spectra} in terms of the specific intensity
change, $\Delta I_\nu$, normalized by
\begin{equation}
  \Delta I_{\rm T0} = {2 h \over c^2} \, \left( {k_{\rm B} T_{\rm rad}
  \over h} \right)^3 \, y \quad , 
\end{equation}
where the Comptonization parameter, 
\begin{equation}
  y = \int n_{\rm e} \sigma_{\rm T} dz \left( {k_{\rm B} T_{\rm e}
      \over m_{\rm e} c^2} \right) \quad ,
\end{equation}
measures the strength of the scattering

The spectrum in Figure~\ref{fig:spectra} has an unusual shape for an
astronomical source. Below about 218~GHz (the frequency of peak
intensity of the microwave background) clusters of galaxies appear as
decrements, while at higher frequencies they are brightness
enhancements. Even weak thermal 
Sunyaev-Zel'dovich effects could be identified against the background
of primary fluctuations by observing over a wide range of
frequencies and making use of this strong spectral signature. This
will be possible with, for example, the data from the {\it Planck}
satellite (Trauber 2001).

At low electron temperatures the spectrum of the thermal
Sunyaev-Zel'dovich effect is independent of $T_{\rm e}$, but
as $k_{\rm B} T_{\rm e}$ rises above about $5 \ \rm keV$, the
increasing fraction of electrons with speeds approaching the speed of
light causes significant spectral modifications from relativistic
effects (Rephaeli 1995b). In the limit of extreme relativistic
electrons the spectrum becomes that of the (inverted) microwave
background itself, since scattered photons are moved to 
energies far outside the microwave band. This nonthermal
Sunyaev-Zel'dovich effect has been discussed as a probe of radio
galaxy electron populations (McKinnon, Owen, \& Eilek 1990), but is
better observed from the scattered photons in the X-ray band than in
the unscattered photons of the radio band (e.g., Hardcastle et
al.~2002). 

The amplitude of the thermal Sunyaev-Zel'dovich effect can be
expressed in terms of the change in total intensity, $\Delta I_\nu$,
at frequency $\nu$, the change in the apparent thermodynamic
temperature of a Planckian spectrum, $\Delta T_\nu$, or the
brightness temperature change
\begin{equation}
 \Delta T_{\rm RJ,\nu} = {c^2 \over 2 k_{\rm B} \nu^2} \Delta I_\nu
                       = f_{\rm T}(\nu) \, T_{\rm rad} \, y \quad , 
 \label{eq:deltat}
\end{equation}
where the spectral function $f_{\rm T}(\nu)$, in the limit of
low $T_{\rm e}$, has the Kompaneets form
\begin{equation}
  f_{\rm T}(\nu) = {x^2 e^x \over \left( e^x - 1 \right)^2} \, 
    \left( x \coth\left({1 \over 2}x\right) - 4 \right) \quad , 
\end{equation}
and
\begin{equation}
 x = {h \nu \over k_{\rm B} T_{\rm rad}}
\end{equation}
is a dimensionless measure of frequency. 

It is clear from Equation \ref{eq:deltat} that the brightness temperature
effect depends only on quantities intrinsic to the cluster, and is
therefore redshift independent. This property of redshift independence
means that clusters of galaxies can be studied in the thermal
Sunyaev-Zel'dovich effect at any redshift where they have substantial
atmospheres. The wide range of redshifts over which the thermal
Sunyaev-Zel'dovich effect can be seen makes the effect an important
probe of cluster evolution.

Practical telescope systems observe a quantity that is a fraction of
the total flux density of a cluster, with the fraction depending
principally on the method of observation and the cluster angular size,
and often being almost constant over a wide range of redshifts
(as in Fig.~\ref{fig:ocrasens}). The total cluster flux density at
frequency $\nu$ is 
\begin{equation}
  \Delta S_\nu = \int \Delta I_\nu \, d\Omega
               \propto {\int n_{\rm e} T_{\rm e} dV \over D_{\rm A}^2} \quad , 
  \label{eq:deltaSnu}
\end{equation}
where the integrals are over the solid angle of the cluster and the
cluster volume. $\Delta S_\nu$ decreases as the inverse square of the
angular diameter distance, $D_{\rm A}$, rather than the inverse square 
of the luminosity distance. This can be thought of as indicating that
Sunyaev-Zel'dovich effect luminosities increase as $(1+z)^4$, because
the luminosity depends on the energy density of the cosmic microwave
background that is available to be scattered. If a given cluster of
galaxies were to be moved from low to high redshift, its flux
density would first decrease, and then increase beyond the redshift of
minimum angular size, or $z = 1.62$ in the $\Lambda$CDM cosmology
adopted here.

Thus at mm wavelengths, the brightest sources in the sky would be
clusters of galaxies if clusters had the same atmospheres in the past
as they have today. That the brightest mm-wave sources are not
clusters does not place a strong constraint on cluster evolution
because $D_{\rm A}(z)$ is a weak function of redshift at $z > 1.6$:
from $z = 1.6$ to 9.8 it decreases by only a factor of $2$.

\subsection{The Kinematic Sunyaev-Zel'dovich Effect}\label{sec:kinematic}
 
If a cluster of galaxies is not at rest in the Hubble flow then the
pattern of radiation illuminating the atmosphere in its rest frame is
anisotropic, and scattering adds a kinematic Sunyaev-Zel'dovich
effect to the thermal effect described in \S~\ref{sec:thermal}
(Sunyaev \& Zel'dovich 1972; Rephaeli \& Lahav 1991). The
kinematic effect has an amplitude that is proportional to
the cluster's peculiar (redshift-direction) velocity, and has a
spectrum that is that of the primary perturbations in the
microwave background radiation (Fig.~\ref{fig:spectra}, lower
panel), where 
%
\begin{equation}
  \Delta I_{\rm K0} = \tau_{\rm e} \, {\upsilon_{\rm z} \over c} \,
                      {2 h \over c^2} \, \left( {k_{\rm B} T_{\rm rad}
                       \over h} \right)^3  \quad .
\end{equation}
The similarity of spectrum makes it difficult to distinguish the
kinematic effect from primary fluctuations in the microwave 
background radiation. This is especially a problem since the kinematic
effect is smaller than the thermal effect by a factor
\begin{equation}
  {\Delta T_{\rm RJ,kinematic} \over \Delta T_{\rm RJ, thermal}}
  = 0.085 \, \left( \upsilon_{\rm z}/1000 \ {\rm km \, s^{-1}} \right) \, 
          \, \left( k_{\rm B} T_{\rm e} / 10 \ {\rm keV} \right)^{-1}
\end{equation}
at low frequencies. Observation near the null of the thermal
Sunyaev-Zel'dovich effect, at $\nu_0 = 218 \ \rm GHz$, gives the
greatest contrast for the kinematic effect, but it is still a small
quantity compared to the confusion from primary structure in the
microwave background radiation on the angular scales of interest (a
few arcmin). This limits the velocity accuracy attainable for any
cluster of galaxies at moderate redshift to about $\pm 150 \ \rm km \,
s^{-1}$, even with perfect data.

\begin{figure*}[t]
\includegraphics[width=1.00\columnwidth,angle=0,clip]{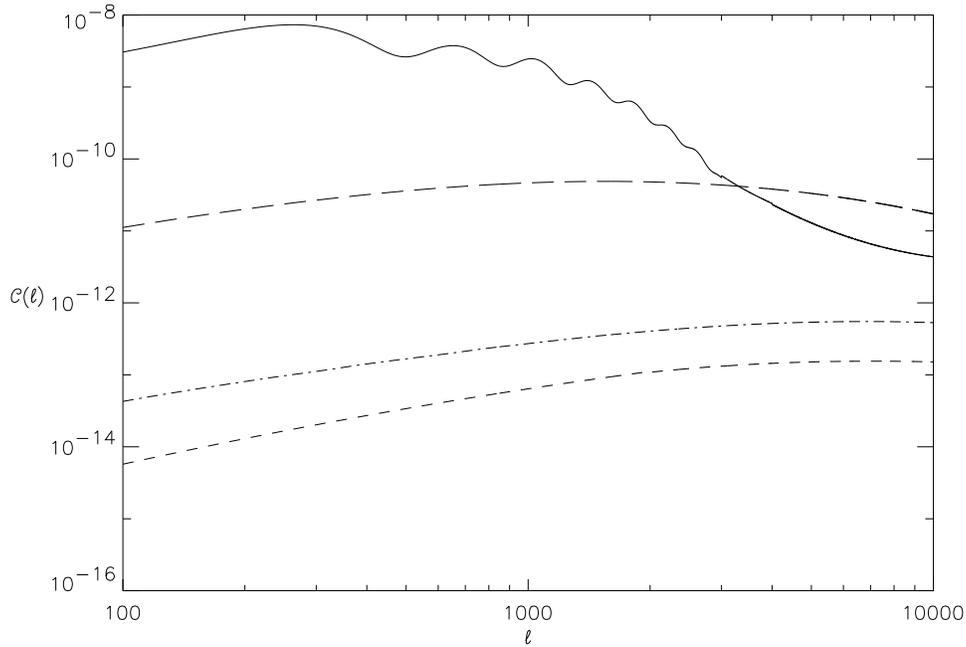}
vskip 0pt \caption{
The lensed primary power spectrum for a $\Lambda$CDM
cosmology with $\Omega_\Lambda = 0.8$, $\Omega_{\rm CDM} = 0.2$, and
$n = -1.4$ (solid line), with the power spectrum expected from the
thermal Sunyaev-Zel'dovich effect (long dashed line), 
kinematic Sunyaev-Zel'dovich effect (dash-dot line), and the 
Rees-Sciama effect from moving clusters (short dashed line) effects.
(From Molnar \& Birkinshaw 2000.)
\label{fig:powerspec}}
\end{figure*}

The relative contributions of the thermal and kinematic
Sunyaev-Zel'dovich effects to the power spectrum of the microwave
background radiation have been calculated by a number of authors
(e.g., da~Silva et al.~2000; Molnar \& Birkinshaw 2000; Springel, White, \& 
Hernquist~2000). The thermal Sunyaev-Zel'dovich effect is expected to
dominate on sufficiently small angular scales
(Fig.~\ref{fig:powerspec}), but on average the kinematic effect makes
a contribution only about $1\%$ that of the thermal effect. Both
effects are larger than the lensing effect of moving clusters of
galaxies. 

\section{Uses of the Sunyaev-Zel'dovich Effect in Cluster Studies}\label{sec:uses}

\subsection{Thermal Energy Content of Clusters}\label{sec:thermalenergy}

The total flux density of the thermal Sunyaev-Zel'dovich effect from a
cluster is given by Equation \ref{eq:deltaSnu}, which can be rewritten
\begin{equation}
  \Delta S_\nu \propto {U_{\rm e} \over D_{\rm A}^2} \quad , 
  \label{eq:flux}
\end{equation}
where $U_{\rm e}$ is the total thermal energy content of electrons in
the intracluster medium. That is, measurements of the redshift
and thermal Sunyaev-Zel'dovich effect of a cluster allow us to infer
the thermal energy content of the intracluster medium without needing
to know the structure of the density or temperature of this gas.

The thermal Sunyaev-Zel'dovich effect therefore can be used as a
calorimeter, measuring the integrated heating to which the cluster gas
has been subjected, although a correction has to be made for energy
lost by (principally X-ray) radiation (Lapi, Cavaliere, \& De~Zotti 2003). If 
the gas is approximately in hydrostatic equilibrium in the cluster, this thermal
energy content should be a good measure of the gravitational 
energy of the cluster, and so a survey for thermal Sunyaev-Zel'dovich
effects should naturally pick out the deepest gravitational potential
wells in the Universe, provided that the ratio of intracluster gas
mass to total mass does not vary too much from cluster to cluster.

\subsection{The Baryonic Mass Content of Clusters}\label{sec:baryonicmass}

If the integrated flux density of the thermal Sunyaev-Zel'dovich
effect and the X-ray spectrum of a cluster are both available, then we
can rewrite Equation \ref{eq:flux} as
\begin{equation}
  \Delta S_\nu \propto \int d\Omega \, \int dz \, n_{\rm e} \, T_{\rm e}
               \propto N_{\rm e} \, \overline{T_{\rm e}} \quad .
\end{equation}
If the mass-weighted mean electron temperature, $\overline{T_{\rm e}}$,
can be approximated by the emission-measure weighted temperature
measured by X-ray spectra, then $\Delta S_\nu$ measures the total
electron count in the intracluster medium, $N_{\rm e}$. A metalicity
measurement for the gas, again from the X-ray spectrum, allows 
$N_{\rm e}$ to be converted into the baryonic mass of the intracluster
medium. This is usually far larger than the stellar 
mass content, and so is a good measure of the total baryonic content
of the cluster. X-ray imaging data can be used to estimate cluster
total masses, using the assumption of hydrostatic equilibrium
(Fabricant, Lecar, \& Gorenstein 1980), and so the baryonic mass
fraction of clusters can be measured. 

Since X-ray images and spectra, and Sunyaev-Zel'dovich effects, can be
measured from massive clusters to $z \approx 1$, a history of the
baryonic mass fraction of clusters can be constructed, although this
is currently biased to the high-mass end of the cluster
population. The result, for the 10--20 clusters for which this has been
done to date (Grego et al.~2001), is that this ratio does not
change significantly with redshift, remaining close to the
result $\Omega_{\rm b}/\Omega_{\rm m} = 0.16 \pm 0.04$ deduced from
the {\it WMAP} results or primordial nucleosynthesis. This suggests that
clusters are fair samples of the mass of the Universe out to the
largest redshifts at which this test has been applied: there is no
strong segregation of dark and baryonic matter during cluster
collapse.

\subsection{Cluster Lensing and the Sunyaev-Zel'dovich
 Effects}\label{sec:lensing} 

It is interesting to consider the possibility of combining
Sunyaev-Zel'dovich effect data with gravitational lensing, rather
than X-ray, data, to study baryonic mass fractions. If the 
ellipticity field $\epsilon_{\rm i}({\bf \theta})$ has been measured
(and corrected to estimate the shear distortion field), then the
surface mass density of a lensing cluster is given by an integral 
\begin{equation}
  \Sigma = -{1 \over \pi} \Sigma_{\rm crit} \, \int d^2 \theta^\prime
  \, K_{\rm i}({\bf \theta}^\prime,{\bf \theta}) \, 
  \epsilon_{\rm i}({\bf \theta}^\prime) \quad , 
\end{equation}
where $\Sigma_{\rm crit}$ is the critical surface density, and the
kernels $K_{\rm i}$ are angular weighting functions. Clearly, the
surface mass density is a linear (though nonlocal) function of the
observable quantity, $\epsilon_{\rm i}({\bf \theta})$. Since the 
Sunyaev-Zel'dovich effect is a linear measure of the
projected baryonic mass density (if the cluster is isothermal), 
the ratio of a Sunyaev-Zel'dovich effect map to a mass map
derived from lensing should give a good measure of the
(projected) radial dependence of baryonic mass fraction. This should
be consistent with the result obtained by applying the standard
assumption of hydrostatic equilibrium to the X-ray data, but is less
susceptible to errors arising from the unknown structure of the
cluster along the line of sight.

\begin{figure*}[t]
\includegraphics[width=1.00\columnwidth,angle=0,clip]{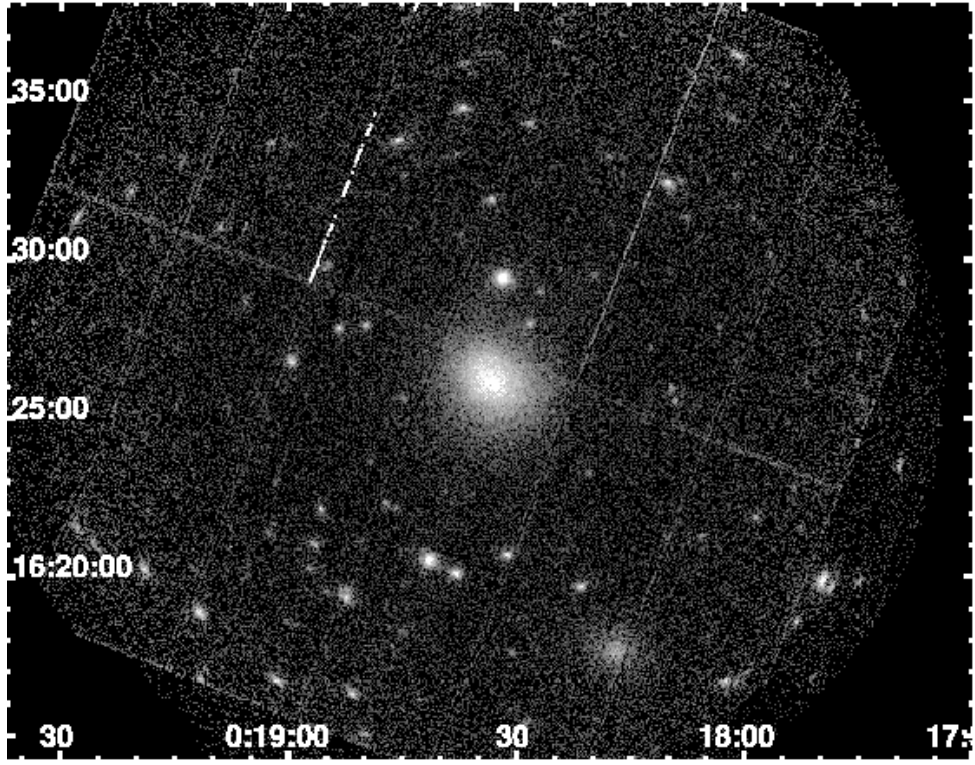}
\vskip 0pt \caption{
A vignetting-corrected $0.3 - 5.0 \ \rm keV$ image of
CL~0016+16, from an {\it XMM-Newton\/} observation (Worrall \&
Birkinshaw 2003). A quasar 3\amin\ north of the cluster, and a
second cluster, 9\amin\ to the south-west, are at similar redshifts
to CL~0016+16. 
\label{fig:0016}}
\end{figure*}

The earliest suggestion of the use of lensing and Sunyaev-Zel'dovich
effect data together to study clusters appears to have been made by
Ostriker \& Vishniac (1986) in the context of the quasar pair 
$1146+111 \rm B,C$, but little work on
the comparison has been done to date, although Dor\'e et al. (2001)
and Zaroubi et al.~(2001) have shown that the addition of lensing data
allows the three-dimensional structure of clusters to be investigated.

A comparison of lensing and X-ray derived masses has
been made for the inner 250~kpc of cluster CL~0016+16
(Fig.~\ref{fig:0016}) by Worrall \& Birkinshaw (2003). Within
this radius, set by the limited lensing data, the masses are
\begin{equation}
  M_{\rm tot} = (2.7 \pm 0.9) \times 10^{14} \ \rm M_\odot 
\end{equation}
and
\begin{equation}
  M_{\rm tot} = (2.0 \pm 0.1) \times 10^{14} \ \rm M_\odot  \quad , 
\end{equation}
found from lensing (Smail et al. 1997, converted to our
cosmology) and X-ray data, respectively. The results agree to within
the limited accuracy of the lensing mass. A combination of these
data with the central Sunyaev-Zel'dovich effect for the cluster,
of $-1.26 \pm 0.07 \ \rm mK$, then leads to a distance-independent
measure of the baryonic mass fraction within the central part of the
cluster of $0.13 \pm 0.02$, close to the value implied by
cosmological parameters.

\subsection{Cluster Structures}\label{sec:structures}

The structures of cluster atmospheres are generally studied from their
X-ray images. The X-ray surface brightness
\begin{equation}
  b_{\rm X} \propto \int n_{\rm e}^2 \, \Lambda(T_{\rm e}) \, dz \quad , 
\end{equation}
where $\Lambda(T_{\rm e})$ is the emissivity function. The X-ray image
can be 
inverted to derive an electron density profile, $n_{\rm e}(r)$, for
the cluster atmosphere under assumptions about the shape of the
atmosphere and the absence of clumping, which would cause
$\overline{n_{\rm e}^2} > \left( \overline{n_{\rm e}}\right)^2$.
Since the Sunyaev-Zel'dovich effects are a projection of $n_{\rm e}$,
rather than the more complicated quantity $n_{\rm e}^2 \Lambda(T_{\rm
e})$, they should give a cleaner measure of the structure of the
atmosphere. However, the restricted angular dynamic range of
measurements of Sunyaev-Zel'dovich effects
(\S~\ref{sec:instruments}), and lower signal-to-noise ratio and angular
resolution of Sunyaev-Zel'dovich effect maps relative to what is
possible with long X-ray exposures, means that X-ray based
structures are superior to those derived entirely from the
Sunyaev-Zel'dovich effects. Furthermore, since the Sunyaev-Zel'dovich
effects are line-of-sight integrals of electron pressure, they are
insensitive to subsonic structures within a cluster, such as cold
fronts, which are easily seen on {\it Chandra\/} X-ray images (e.g.,
Markevitch et al.~2000). 

The different $n_{\rm e}$ dependencies of the Sunyaev-Zel'dovich effect
and X-ray surface brightness also implies that clusters of galaxies
have larger angular sizes in Sunyaev-Zel'dovich effects than in the
X-ray, and that the X-ray emission is more sensitive to cluster cores,
while the Sunyaev-Zel'dovich effects are more weighted toward the low-density 
envelopes. However, the low surface brightness of cluster envelopes,
and low signal-to-noise ratio even in the centers of most Sunyaev-Zel'dovich
effect maps, means that the X-ray data are superior out to the largest
radii to which gas has been detected.

\subsection{Cluster Distances}\label{sec:distances}

Perhaps the most widely discussed use of the thermal
Sunyaev-Zel'dovich effect has been to measure the distances of
clusters of galaxies. In its simplest form, the method is to compare
the X-ray surface brightness of a cluster on some fiducial line of
sight, 
\begin{equation}
 b_{\rm X0} \propto n_{\rm e0}^2 \, \Lambda(T_{\rm e0}) \, L \quad , 
\end{equation}
with the thermal Sunyaev-Zel'dovich effect on the same line of sight, 
\begin{equation}
 \Delta T_{\rm RJ,0} \propto n_{\rm e0} \, T_{\rm e0} \, L \quad ,
\end{equation}
and eliminate the scale (perhaps central) electron density in the
cluster and find an absolute measurement of the path length along some
fiducial line of sight
\begin{equation}
  L \propto {\Delta T_{\rm RJ,0}^2 \over b_{\rm X0}} \cdot {\Lambda(T_{\rm
     e0}) \over T_{\rm e0}^2}
  \quad .
\end{equation}
This path length can then be compared with the angular size of the cluster
to infer the cluster's angular diameter distance, under the assumption
that the cluster is spherical. This technique has been used for a
number of clusters (e.g., Hughes \& Birkinshaw 1998; Mason, Myers, \&
Readhead 2001; Reese et al.~2002). A recent recalculation of the
distance to CL~0016+16 (Fig.~\ref{fig:0016}) using this technique gave
$D_{\rm A} = 1.16 \pm 0.15 \ \rm Gpc$ (Worrall \& Birkinshaw 2003),
and implies a Hubble constant of $68 \pm 8 \pm 18 \ \rm km \, s^{-1} \,
Mpc^{-1}$.

While single cluster distances are likely to be error-prone, because
of their unknown three-dimensional shapes (and hence the large
systematic error on the result for CL~0016+16), a suitably chosen
sample of clusters, without an orientation bias, can be used to map
the Hubble flow to $z \approx 1$, and hence to measure a number of
cosmological parameters. Molnar, Birkinshaw, \& Mushotzky (2002) have
shown that a set of about 70~clusters could provide useful measure of the
equation-of-state parameter, $w$, as well as the Hubble
constant. A recent review of the state of measurement of $D_{\rm
A}(z)$ using this technique is given by Carlstrom et al. (2002).

It is critical that the absolute
calibrations of the X-ray and Sunyaev-Zel'dovich effect data are
excellent, and that cluster substructure is well modeled, if
distances are to be estimated in this way. Since
clusters are relatively young structures, and likely to be changing
significantly with redshift, variations in the amount of substructure
with redshift might be a significant source of systematic
error. High-quality X-ray imaging and spectroscopy are therefore 
essential if the distances obtained are to be reliable.

Finally, a crucial element of this technique is that the set of
clusters used should be free from any biasing selection
effect. Perhaps the most important such effect is that of 
orientation: if clusters are selected by any surface brightness
sensitive criterion, then they will tend to be preferentially aligned
relative to the line of sight. Since the distance measurement
technique relies on comparing the line-of-sight depth of the cluster,
$L$, with the cross line-of-sight angular size, $\theta$, such a
selection will inevitably induce a bias into the measured
$D_{\rm A}(z)$ function.

Clumping of the intracluster medium and a number of other problems can
also cause biases in the results. Comprehensive reviews of such
biases can be found in most papers on Hubble constant measurements
using this technique, and in Birkinshaw (1999).

\subsection{Cluster Peculiar Velocities}\label{sec:velocities}

The kinematic Sunyaev-Zel'dovich effect (\S~\ref{sec:kinematic}) is
heavily confused by primary anisotropies in the microwave background
radiation since cluster peculiar velocities are expected to be less
than $100 \ \rm km \, s^{-1}$. It seems likely, therefore, that
measurement of the kinematic effect will only be possible on a
statistical basis, by comparing the brightness of the microwave
background radiation toward the clusters of galaxies with positions
in the field. Since the kinematic effect, like the thermal effect, is
redshift independent, such a statistical measurement would be an
important check on the changing velocities of clusters with redshift
as structure develops. Efforts to measure the kinematic effect
continue, although only relatively poor limits on cluster velocities,
of order $1000 \ \rm km \, s^{-1}$, have been obtained to date
(Holzapfel et al.~1997a; LaRoque et al.~2003b).

\subsection{Cluster Samples}\label{sec:detection}

Since the Sunyaev-Zel'dovich effects are redshift-independent in
$\Delta T_{\rm RJ}$ terms, and the observable $\Delta S_\nu$ are
almost redshift-independent, it follows that Sunyaev-Zel'dovich effect
surveys should be more effective than X-ray or optical surveys at
detecting clusters of galaxies at large redshift. Such samples of
clusters, selected by their Sunyaev-Zel'dovich effects, will be
unusually powerful in being almost mass limited, and counts of the
number of such clusters as a function of redshift provide a good
method of measuring $\sigma_8$ (Fan \& Chiueh 2001).

Such samples are also ideal for work on the distance scale, and hence
the measurement of the equation-of-state parameter, $w$.

\section{Instruments and Techniques}\label{sec:instruments}

Many observations of the thermal Sunyaev-Zel'dovich effect have
followed its first reliable detection, in Abell~2218 (Birkinshaw et al.
1978). While the earliest observations were made with
single-dish radiometers, more recently work has been done with 
interferometers and bolometer arrays. The quality of the observational
data is advancing rapidly, so that the review of the observational
situation that I wrote in 1999 (Birkinshaw 1999) grossly
underrepresents the number of good detections and images of the effect.

The most effective observations of the Sunyaev-Zel'dovich effects
over the past few years have used interferometers, notably the Ryle
telescope (Jones et al.~1993) and the BIMA and OVRO arrays (Carlstrom,
Joy, \& Grego 1996). Since the thermal Sunyaev-Zel'dovich effects of
rich clusters typically have angular sizes exceeding 1\amin, 
most of the correlated signal appears on the shortest ($< 2000\lambda$)
baselines. These baselines are usually underrepresented in the
interferometers, so that all these instruments had to be retrofitted
by increasing the observing wavelength, $\lambda$, or altering the array
geometry. The effectiveness of the changes is demonstrated by the
large number of high signal-to-noise ratio detections of clusters of galaxies
that interferometers have produced (e.g., Joy et al.~2001; Cotter et al.~2002;
Grainge et al.~2002). 

The major advantage of interferometers is their ability to reject
contaminating signals from their surroundings and the atmosphere, and
to spatially filter the data to exclude radio point sources in the
fields of the clusters. Extremely long interferometric integrations
are possible before parasitic signals degrade the data. The
limitation, at present, is that the retrofitted arrays 
are not ideally matched to the purpose of mapping the
Sunyaev-Zel'dovich effect. This limitation is being addressed by the
construction of a new generation of interferometers, including AMI
(Kneissl et al.~2001), SZA (Mohr et al.~2002), and AMiBA (Lo et
al.~2001), which will have enough sensitivity to undertake deep
blank-field surveys for clusters at $z > 1$.

Single-dish radiometer systems, particularly when equipped with radiometer 
arrays, are efficient for finding strong Sunyaev-Zel'dovich effects. Their large
filled apertures can integrate the signal over much of the solid angle
of a cluster. However, spillover and residual atmospheric noise limit
the length of useful integrations and hence the sensitivity that can
be achieved (Birkinshaw \& Gull 1984). Since practical systems always
involve differencing between on-source and off-source sky regions,
this technique, just as interferometry, is sensitive only to 
clusters smaller than some maximum angular size, and hence at
sufficiently large redshift. Finally, there is 
no simple way of removing the signals of contaminating (and often
variable) radio sources that appear superimposed on the
Sunyaev-Zel'dovich effect. This further restricts the set of clusters 
that can be observed effectively.

Modern antennas provide new opportunities for radiometer systems since
they have superior spillover characteristics and can be equipped with
radiometer arrays. The 100-meter Green Bank Telescope is an obvious 
example, and the OCRA project (\S~\ref{sec:newprojects}) is intending 
to provide a fast survey capability for this and other large telescopes. 

Bolometers are capable of measuring the spectrum of the Sunyaev-Zel'dovich 
effect above $\sim 90$~GHz, and could separate the thermal Sunyaev-Zel'dovich
effect from the associated kinematic effect or from primary anisotropy
confusion. The intrinsic sensitivity of bolometers should allow 
fast measurements of targeted clusters, or high-speed surveys of large
regions of sky. However, bolometers are exposed to a high level of
atmospheric and other environmental signals, and so the differencing
scheme used to extract sky signals from the noise must be 
of high quality.

There have been two attempts to use the spectral capability of
bolometers to determine cluster peculiar velocities
(Holzapfel et al. 1997a; LaRoque et al. 2003b), and a similar
experiment can use the spectral distortion of the thermal effect to 
measure the temperature of the microwave background radiation in
remote parts of the Universe and test whether $T_{\rm rad} \propto
(1+z)$ (Battistelli et al. 2002). Newer bolometer arrays, such as
BOLOCAM (Glenn et al. 1998) and ACBAR (Romer et al. 2001),
should allow fast surveys for clusters and could make
confusion-limited measurements of the cluster velocity if the flux 
scale near 1~mm is well calibrated.

\subsection{Recent Advances}\label{sec:advances}

A major advance over the past few years has been the change from
making and reporting Sunyaev-Zel'dovich effects for individual
clusters to reporting for samples of clusters. The samples most
favored are based on X-ray surveys (e.g., Mason et
al.~2001; LaRoque et al. 2003a), and can be assumed to be
orientation-independent provided that the selection is at X-ray fluxes
far above the sensitivity limit of the survey. The combination of good
X-ray and Sunyaev-Zel'dovich effect data is leading to reliable
results for baryon fractions and the Hubble constant.

\begin{figure*}[t]
\includegraphics[width=1.00\columnwidth,angle=0,clip]{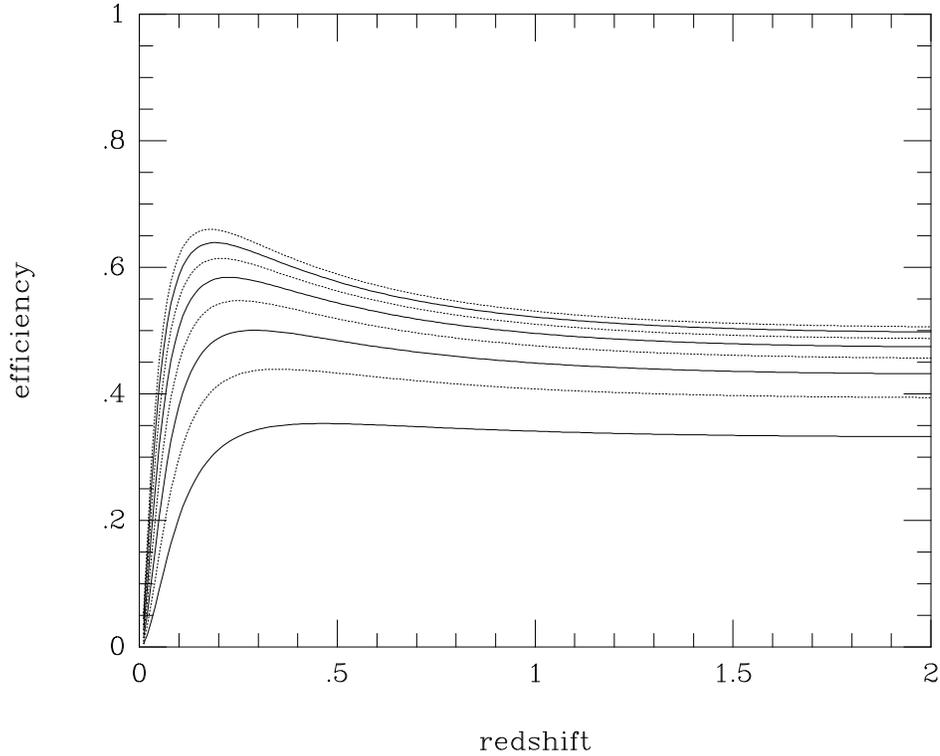}
\vskip 0pt \caption{
The relative sensitivity of observations of the thermal
Sunyaev-Zel'dovich effect in a cluster of galaxies as a function of
differencing angle (from 1\farcm5 to 5\farcm0 from the bottom to
the top curve) and redshift for the
One Centimeter Radiometer Array (OCRA; Browne et al.~2000). Note the
flatness of the sensitivity function at $z > 0.5$ for any
differencing angle.
\label{fig:ocrasens}}
\end{figure*}

The next improvement will clearly be to obtain samples of clusters
selected entirely in the Sunyaev-Zel'dovich effect, on the basis of
blind surveys. Such surveys require sensitive, arcminute-scale, instruments
capable of covering many $\rm deg^2$ of sky in a reasonable length of
time --- a good first target would be to cover $10 \ \rm deg^2$ to a
sensitivity limit $\Delta T_{\rm RJ} \approx 100 \ \rm \mu K$, yielding a
sample of tens of clusters selected only through their
Sunyaev-Zel'dovich effects. Such cluster samples would be ideal for
cosmological purposes, since they would be almost mass limited. Since
the observable thermal Sunyaev-Zel'dovich effect depends linearly on
the properties of clusters, and on the angular diameter distance,
which is a slow function of redshift for $z \approx 1$, a cluster of a
given mass can be detected with almost the same efficiency at any
redshift $> 0.5$. This is illustrated for the OCRA instrument (Browne
et al.~2000) in Figure~\ref{fig:ocrasens}.

\begin{figure*}[t]
\includegraphics[width=1.00\columnwidth,angle=0,clip]{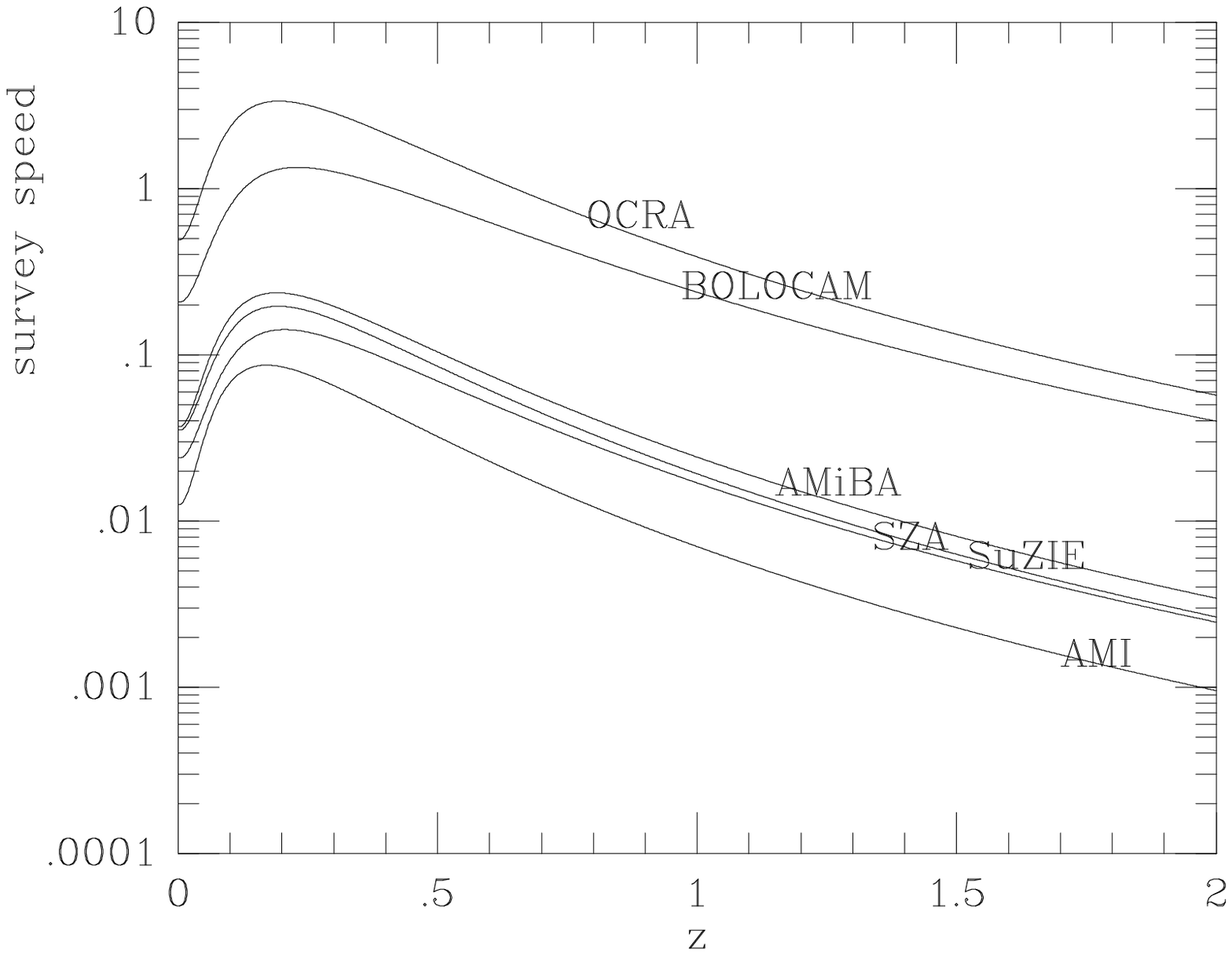}
\vskip 0pt \caption{
The relative speeds of several future Sunyaev-Zel'dovich
effect instruments, as a function of redshift, for a model in which
the cluster gas evolves in a non self-similar fashion. The
experiments shown are OCRA (Browne et al.~2000;
\S~\ref{sec:newprojects}), BOLOCAM
(Glenn et al.~1998), AMiBA (Lo et al.~2001;
\S~\ref{sec:newprojects}), SZA (Mohr et al.~2002), SuZIE
(Holzapfel et al. 1997b), and AMI (Kneissl et al.~2001).
\label{fig:speeds}}
\end{figure*}

Many of the new projects are aimed at blind surveys of
the microwave background sky to find Sunyaev-Zel'dovich effects, and
hence complete cluster samples. Figure~\ref{fig:speeds} compares the
survey speeds of these projects, ranging from radiometer and
bolometer arrays (OCRA, BOLOCAM) to interferometers (AMiBA, SZA,
AMI). AMiBA and OCRA are discussed in more detail in the next
section. 

\subsection{Two New Projects: AMiBA and OCRA}\label{sec:newprojects}

AMiBA, the Array for Microwave Background Anisotropy, is described by
Lo et al.~(2001). It is a platform-based system, operating at 95~GHz
with a bandwidth of 20~GHz, with up to 19 1.2-m antennas, providing an
11\amin\ field of view and approximately 2\amin\ angular
resolution. Its flux density sensitivity, 1.3~mJy per beam in
1~hour, makes it a highly effective interferometer for surveys: a
3-hour observation suffices to detect a $z = 0.5$ cluster with mass
$2.7 \times 10^{14} \ \rm M_\odot$ at $5\sigma$, and clusters with
about a third of this mass could be detected well above the confusion
limit in longer integrations.

A trial platform carrying two smaller antennas has been operating on
Mauna Loa for some months, and various aspects of the system are being
tried out and debugged. AMiBA should be operational in the 2003/4
winter season for studies of primary structure in the microwave
background radiation, and full Sunyaev-Zel'dovich effect operations
should start about 1~year later. 

The high density of short antenna-antenna spacings and the 3-mm
operating wavelength make AMiBA the most sensitive of the planned
interferometers. However, AMI at 15~GHz (Kneissl et al.~2001), the SZA
at 30~GHz (Mohr et al.~2002), and AMiBA at 95~GHz have usefully
complementary angular resolutions and operating frequencies. Although
the spread of frequencies provided by these three systems is
inadequate for useful spectral work, to separate primary anisotropies and
kinematic effects from the thermal signal, it is sufficient to provide
a control against nonthermal radio sources, and an important
cross-check on cluster counts.

Several thousand individual pointings with AMiBA would be needed for
a shallow survey of the entire 64~$\rm deg^2$ \textit{XMM-Newton} survey
field (Pierre et al. 2001). While most of the $\sim 10^3$ Sunyaev-Zel'dovich 
effect detections expected will correspond to X-ray clusters, a comparison of
the \textit{XMM-Newton} and AMiBA detection functions shows that AMiBA
will be better at detecting clusters beyond $z = 0.7$, and could
find clusters with X-ray luminosities $> 4 \times 10^{43} \ \rm erg \,
s^{-1}$ (in $0.5-10 \ \rm  keV$) to $z > 2$ for later, deeper, X-ray
and infrared follow-up. 

While AMiBA has the interferometric advantages of rejecting
contaminating environmental signals, high sensitivity and fast survey
speed are better achieved by radiometer arrays. OCRA, the One
Centimeter Radiometer Array (Browne et al.~2000), has been designed to
make use of recent advances in radiometer technology to perform 
wide-field surveys for Sunyaev-Zel'dovich effects and radio sources.

OCRA is an array of $\sim 100$ 30-GHz radiometers, based on {\it Planck}
technology, housed in a single cryostat at the secondary (or
tertiary) focal plane of a large antenna, and providing 1\amin\ beams
with $\sim 3$\amin\ beam separations. Each radiometer has excellent
flux density sensitivity at 30~GHz. This is a large and expensive
system, and two preliminary (but scientifically useful) versions of
OCRA will be used before OCRA is built.

The prototype, OCRA-p, is a two-beam system now being tested
on the Torun 32-m radio telescope. With a system temperature 
below $40 \ \rm K$, we expect to achieve a flux density sensitivity of
5~mJy in 10~s provided that the atmospheric noise is well controlled.
OCRA-p will be followed by a FARADAY project receiver with eight
beams, which uses MMICs on InP substrates, and which is funded by a
European Union grant. This receiver is currently under construction
and will be in use in about a year. It will achieve a significant
improvement in mapping speed over OCRA-p, though still be a factor
$\sim 10$ slower than OCRA itself.

OCRA will be an extremely fast instrument for finding high-redshift
clusters. A comparison of its mapping speed with some other systems
is shown in Figure~\ref{fig:speeds}. While OCRA is more susceptible to
radio source confusion than AMiBA (indeed, the study of the radio
source population at 30~GHz is one aim of OCRA), recent work on the
probability of detecting Sunyaev-Zel'dovich effects from clusters at 18.5~GHz 
(Birkinshaw, in preparation) suggests that the level of source confusion will 
not prevent OCRA from detecting most rich, distant clusters. Even the
FARADAY receiver will be a highly competitive survey instrument.

\section{Summary}\label{sec:summary}

The thermal Sunyaev-Zel'dovich effects of clusters of galaxies, expressed in
brightness temperature terms, are redshift-independent measures of the
thermal energy content of clusters: they are, effectively,
calorimeters of energy releases in clusters after account is taken of 
the energy radiated by line and bremsstrahlung emission (principally
in the X-ray band). With the assistance of X-ray and/or gravitational
lensing data, the thermal Sunyaev-Zel'dovich effect measures a wide
range of cluster properties, including distance.

Thermal Sunyaev-Zel'dovich effects are mass finders, with strong
associations with rich clusters of galaxies, and should be good probes
of the most massive structures that exist in the Universe at any
redshift. Deep Sunyaev-Zel'dovich effect studies are therefore an
excellent way of discovering the degree of cluster formation in the
Universe to high redshift, and counts of clusters from forthcoming
surveys should place strong constraints on the processes of cluster
formation. 

Two weaker Sunyaev-Zel'dovich effects, the kinematic and polarization
effects, are also of interest, but are harder to measure. Both can
give information about the speeds of clusters of galaxies, and hence
the evolving dynamics of gravitating objects in the Universe, but
both are subject to considerable confusion from primary structures
in the microwave background radiation and are therefore likely to be
detectable only in a statistical sense for populations of clusters.

Over the past 20 years, the thermal Sunyaev-Zel'dovich effect has
gone from being a curiosity to a major tool for cosmology and cluster
physics. Substantial results based on Sunyaev-Zel'dovich effect work
are to be expected within the next 10 years.

\end{document}